\documentclass[preprint,showpacs,amsmath,amssymb,aps]{revtex4}
\usepackage{graphicx}
\usepackage{amssymb}
\usepackage{epstopdf}
\usepackage{dcolumn}
\usepackage{bm}
\usepackage{subfigure}

\begin{document}
\title{A novel wear-resistant magnetic thin film material based on a Ti$_{1-\text{x}}$Fe$_{\text{x}}$C$_{1-\text{y}}$ nanocomposite alloy}
\author{Stojanka Bijelovic$^{1}$, Mikael R{\aa}sander$^{2}$, Ola Wilhelmsson$^{3}$, Erik Lewin$^{3}$, Biplab Sanyal$^{2}$, Ulf Jansson$^{3}$, Olle Eriksson$^{2}$ and Peter Svedlindh$^{1}$}

\affiliation{$^{1}$Department of Engineering Science, Uppsala University, P.O. Box 534, SE-751 21 Uppsala, Sweden}

\affiliation{$^{2}$Department of Physics and Materials Science, Uppsala University, P.O. Box 530, SE-751 21 Uppsala, Sweden}

\affiliation{$^{3}$ Department of Materials Chemistry, Uppsala University, P.O. Box 538, SE-751 21 Uppsala, Sweden}

\date{\today}

\begin{abstract}
In this study we report on the film growth and characterization of thin (approximately 50 nm thick) Ti-Fe-C films deposited on amorphous quartz. The experimental studies have been complemented by first principles density functional theory (DFT) calculations. Upon annealing of as-prepared films, the composition of the metastable Ti-Fe-C film changes. An iron-rich phase is first formed close to the film surface, but with increasing annealing time this phase is gradually displaced toward the film-substrate interface where its position stabilizes. Both the magnetic ordering temperature and the saturation magnetization changes significantly upon annealing. The DFT calculations show that the critical temperature and the magnetic moment both increase with increasing Fe and C-vacancy concentration. The formation of the metastable iron-rich Ti-Fe-C compound is reflected in the strong increase of the magnetic ordering temperature. Eventually, after enough annealing time ($\geq 10$ minutes), nano-crystalline $\alpha$-Fe starts to precipitate and the amount and size of these precipitates can be controlled by the annealing procedure; after 20 minutes of annealing, the experimental results indicate a nano-crystalline iron-film embedded in a wear resistant TiC compound. This conclusion is further supported by transmission electron microscopy studies on epitaxial Ti-Fe-C films deposited on single crystalline MgO substrates where, upon annealing, an iron film embedded in TiC is formed. Our results suggest that annealing of metastable Ti-Fe-C films can be used as an efficient way of creating a wear-resistant magnetic thin film material.
\end{abstract}
\maketitle
\section{\label{sec:intro}Introduction}

Thin films of the transition metal carbides (TMC) have received much attention due to their high wear-resistance, low friction and high chemical stability \cite{Hauert,Cavalerio, Martinez}. A very interesting application for this type of material is wear-resistant, low-friction magnetic coatings. Such materials could be used in e.g. sensors and electronics. A problem with carbide-based nanocomposites is that the magnetic elements Fe, Ni and Co are weak carbide-formers and in fact only form metastable carbides although they have a high activation energy towards decomposition into graphite and metal.  Furthermore,  late 3$d$ transition metal carbides are less hard and wear-resistant than the early 3$d$ transition metal carbides.  A possible solution to this problem could be to form a solid solution of a magnetic element in a stable transition metal carbide such as TiC. Such solid solutions may retain most of the mechanical and tribological properties (see e.g. \cite{Cavalerio,Hono,Hasegawa}) and also be tuned with respect to magnetic properties. A problem with this approach is that the solid solubility of e.g. Fe and  Ni in TiC at equilibrium is very low ($<1$ at$\%$) \cite{Vilars}. However, it is well-known that some thin film deposition processes such as magnetron sputtering at low substrate temperature can give films with very high solid solubility. For example, Wilhelmsson \textit{et al}. have previously demonstrated that as much as 50$\%$ of the Ti in TiC can be substituted for Fe \cite{Ola1}. During annealing the metastable material decomposes to TiC and metallic precipitates. It was  demonstrated that this concept can be used to produce wear-resistant magnetic films but no details of the influence of e.g. temperature on the magnetic properties have yet been published.  Furthermore, Trindade and Tome \textit{et al}. have previously reported on the synthesis of thin films within the Ti-Fe-C system  but the magnetic properties of this material were not measured \cite{ref5,ref6}.

The aim of the present study is to further investigate how the magnetic properties of a Ti$_{1-\text{x}}$Fe$_{\text{x}}$C$_{1-\text{y}}$ film is influenced by annealing. The films have been characterized by x-ray diffraction (XRD), x-ray photoelectron spectroscopy (XPS) and superconducting quantum interference device (SQUID) magnetometry. Selected films have also been investigated by transmission electron microscopy (TEM).  The experimental work is supported by first principles density functional theory (DFT) calculations, yielding information on mixing enthalpies, exchange parameters and magnetic ordering temperatures.

\section{\label{sec:Experi}Experimental} 

Thin films of Ti$_{1-\text{x}}$Fe$_{\text{x}}$C$_{1-\text{y}}$ with nominal composition $x = 0.4$ and $y = 0.5$ were deposited by non-reactive dc magnetron sputtering in an Ar-discharge of 3 mTorr (base pressure of $<$10$^{-9}$ Torr). The anticipated composition was achieved by individual control of the magnetron current of the three 2"-elemental targets of Ti, Fe and C; the purity of each target was 99.995\%, 99.99\% and 99.995\%, respectively. The substrate holder was held at a floating potential of about -1V and was rotating during film growth. The films were $\approx$50 nm thick and deposited on amorphous quartz. Prior to deposition the substrates were ultra-sonically degreased in acetone and isopropanol for 5 minutes each. Thereafter, they were pre-heated in the deposition chamber for 30 minutes at 450$^\text{o}$C followed by film growth at the same temperature.  The films were post-heat treated in a high-vacuum furnace (10$^{-7}$ Torr) at 650$^\text{o}$C for annealing times of 2, 5, 8, 10 and 20 minutes.

The epitaxial (Ti$_{1-\text{x}}$Fe$_{\text{x}})$C$_{1-\text{y}}$ films with nominal composition $x = 0.3$ and $y = 0.7$ were fabricated by the same process and post-heat treated at 850$^\text{o}$C for an annealing time of 60 minutes. These films were $\approx 50$ nm thick and deposited on single crystalline MgO(100) substrates.

The chemical composition and elemental-chemical state of the deposited films were analysed by x-ray photoelectron spectroscopy (XPS) employing a PHI Quantum 2000 with monochromatized AlK$\alpha$ radiation. Compositional depth profiles were achieved by sequential sputtering using 4keV Ar$^+$. The crystal structure and crystallinity were studied by x-ray diffraction (XRD) using a Philips X'pert equipment (radiation CuK$\alpha$) with a mirror on the primary side and a parallel collimator 0.27$^\text{o}$ on the secondary side. The films deposited on amorphous quartz were analysed by grazing incident (GI) instrumental scan geometry employing an incident angle of 0.5$^\text{o}$. 

The epitaxial films were also examined in a transmission electron microscope (TEM) equipped with an energy-filter system, which was used to record inelastically scattered electrons resulting in energy-filtered images (EF-TEM). For this purpose a Gatan image filter, GIF2002, was used containing a magnetic prism that bends the electrons 90$^\text{o}$ off the initial optical axis. Electrons losing different amount of energy are separated into an energy-dispersive spectrum. For elemental mapping an energy selecting slit was used for picking out a certain energy range.

Magnetization measurements were performed in a Quantum Design MPMS-XL Superconducting Quantum Interference Device (SQUID) magnetometer. Magnetization ($M$) versus temperature ($T$) was studied between 5 K and 400 K following a field cooled protocol. A weak magnetic field ($H = 50$ Oe) was applied at 400 K and the magnetization was measured as the sample cooled down to 5 K. Magnetization versus field measurements were performed at 10 K between $10^4$ Oe and $-10^4$ Oe.

\section{\label{sec:theory}Theory}
First principles density functional theory calculations (DFT) \cite{dft1,dft2} using the Green's function Kohn-Korringa-Rostoker method (KKR) while employing the atomic sphere approximation (ASA) \cite{KKR1,KKR2},  has been used in order to calculate total energies, magnetic moments, exchange parameters and electronic structure of a number of ternary Ti$_{1-\text{x}}$Fe$_{_{\text{x}}}$C$_{1-\text{y}}$-phases with varying composition of Ti, Fe and C. The local spin-density approximation (LSDA) \cite{PBELDA96} was used for the exchange-correlation functional in all calculations.

Binary TiC crystallizes in the B1 (or NaCl) structure which consists of two intersecting face centered cubic lattices (one for Ti and one for C) shifted half-way along the body diagonal relative to each other. In order to properly fulfill the requirements for the ASA, we have introduced two empty spheres along the body diagonal of the structure \cite{Korzhavyi}. Earlier studies have shown that other metals in TiC substitutes for Ti and results in a two component metal lattice. If the other metal is a poor carbide former, such as Fe, this substitution in turn reduces the otherwise strong covalent bond between Ti and C atoms, yielding a ternary system with a lower C content than normal TiC. It is therefore necessary to calculate the magnetic properties for varying carbon content, since a large number of C vacancies are present on the C-lattice.

The substitutional disorder of Fe-atoms and C-vacancies were treated within  the coherent potential approximation (CPA). The CPA is a single-site approximation and local environment effects are therefore neglected. These effects may have a large contribution due to local interactions and relaxations \cite{localeff}. However, even though the CPA neglects local effects, the favorable properties when it comes to computational effort compared to supercell methods makes it very efficient.

The magnetic properties have been evaluated by first calculating the Heisenberg exchange parameters $J_{ij}$ between a pair of magnetic atoms positioned at $\mathbf{R}_{i}$ and $\mathbf{R}_{j}$ according to the theory by Liechtenstein \textsl{et al}. \cite{Liechtenstein} employing the magnetic force theorem, where the exchange parameters are given by
\begin{equation}\label{eq:jij}
J_{ij} = \frac{1}{4\pi}\int^{E_{F}}\,dE\,\text{Im}\left\{Tr_{L}(\Delta_{i}T_{\uparrow}^{ij}\Delta_{j}T_{\downarrow}^{ij})\right\},
\end{equation}
where $\Delta_{i}=t_{i\uparrow}^{-1}-t_{i\downarrow}^{-1}$, $t$ being the on-site scattering matrix, and $T$ the scattering matrix operator related to the off-diagonal elements of the Green's function. The trace in the above equation has been performed over the orbital indices of the scattering matrices. A positive (negative) $J_{ij}$ represent a ferromagnetic (anti-ferromagnetic) interaction.

The Curie temperature has been calculated using both the mean field approximation (MFA) and Monte Carlo (MC) calculations.
The MFA Curie temperature is evaluated according to
\begin{equation}
T_{c}^{MFA} = \frac{2x}{3k_{B}}\sum_{j\neq0}J_{0j},
\end{equation}
where $x$ is the concentration of magnetic atoms and $k_{B}$ is Boltzmann's constant. Generally, the MFA is known to overestimate critical temperatures, especially for low concentrations of magnetic atoms.

The MC calculations have been performed using the classical zero field Heisenberg Hamiltonian
\begin{equation}
H = -\sum_{i\neq j}J_{ij}\mathbf{e}_{i}\cdot\mathbf{e}_{j},
\end{equation}
where the magnetic exchange parameters are given by Eq. (\ref{eq:jij}) and $\mathbf{e}_{i}$ is the unit vector of the magnetic moment at position $\mathbf{R}_{i}$. The simulations were performed using the single spin-flip Metropolis algorithm for varying sizes of the simulation cell and the critical temperature was obtained from the crossing of the fourth order cumulant of the magnetization for different sizes of the simulation cell \cite{Landaubinder}.

\section{\label{sec:Results}Results}
\subsection{\label{sec:XPS}Phase and microstructure characterization}

\begin{figure}
\begin{center}
\includegraphics[height=6cm]{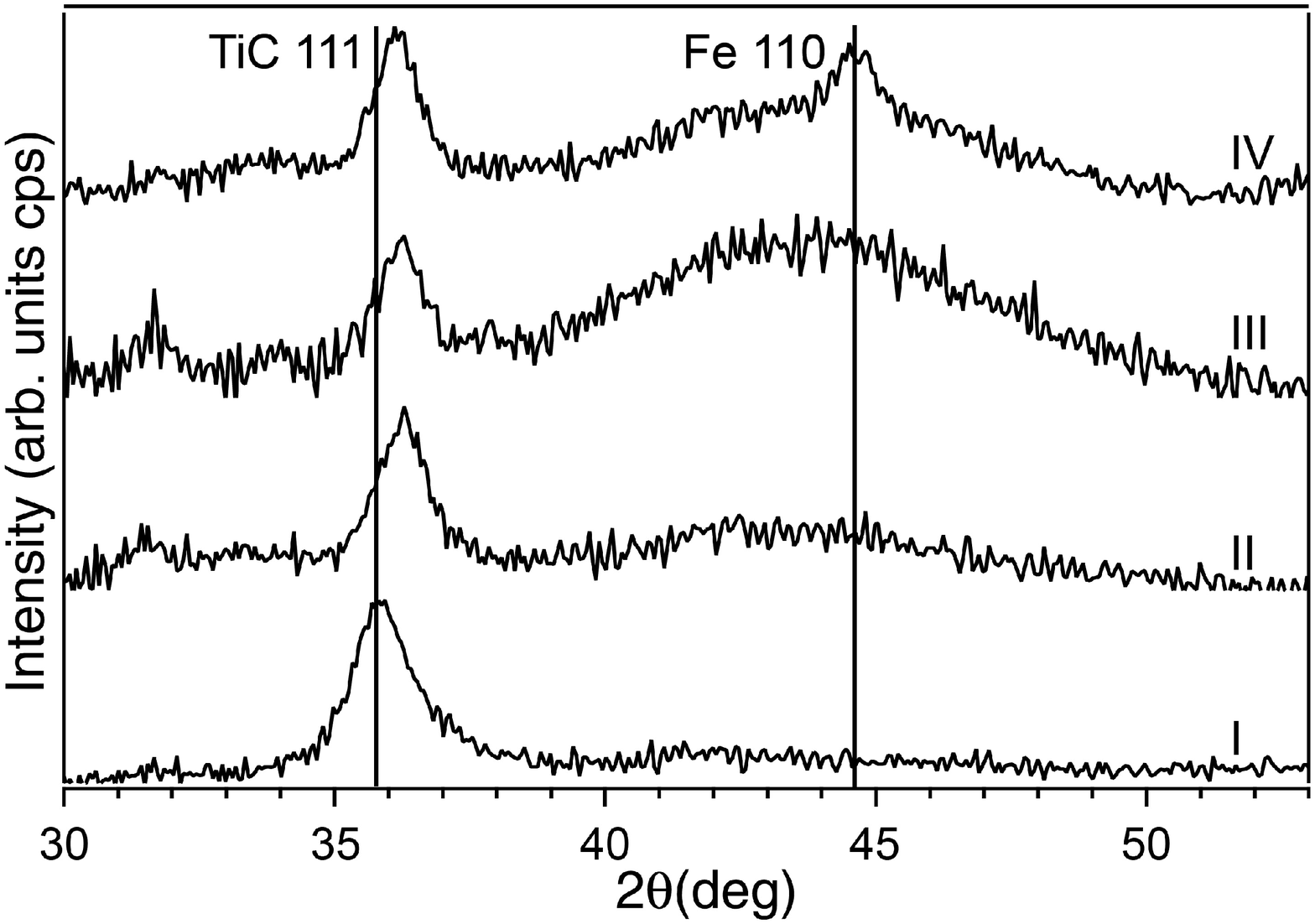}
\caption{X-ray diffractograms of grazing incidence scans for the as-deposited  film and films heat treated at 650$^\text{o}$C. I, II, III and IV label the annealing times of 0, 5, 10 and 20 minutes, respectively. After 20 minutes of heat treatment a peak attributed to $\alpha$-Fe is seen.}
\label{fig:XRD}
\end{center}
\end{figure} 

\begin{figure}
\begin{center}
\includegraphics[height=6cm]{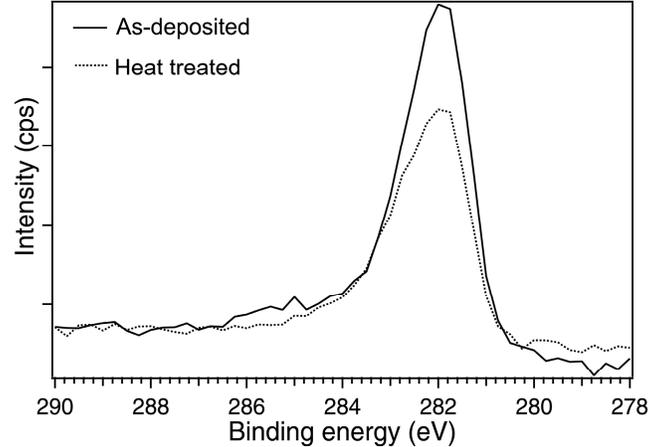}
\caption{C1s spectrum after a short Ar$^{+}$-sputter cleaning of the surface for the as-prepared film (solid line) and for a film annealed for 20 minutes at 650$^\text{o}$C (dashed line). All heat treated films give the same result.}
\label{fig:Cs1}
\end{center}
\end{figure} 

Figures \ref{fig:XRD} and \ref{fig:Cs1} show GI X-ray diffractograms and C1s XPS spectra of an as-deposited and an annealed film with the total composition Ti$_{0.6}$Fe$_{0.4}$C$_{0.5}$. In an earlier study such as-deposited films have been characterized in detail as a function of composition \cite{Ola1}. An important result was that Fe was dissolved into the carbide structure by a substitutional solid solution with Fe on the Ti sites. Furthermore,  it was found that sputtered Ti$_{1-\text{x}}$Fe$_{\text{x}}$C$_{1-y}$ films with $y \geq 0.5$ consist of nanocrystalline (Ti,Fe)C grains while higher carbon contents yield a nanocomposite with nanocrystalline (Ti,Fe)C grains embedded in an amorphous carbon matrix. The results in Figs. \ref{fig:XRD} and \ref{fig:Cs1} confirm these earlier results. The diffractogram I in Fig. \ref{fig:XRD} exhibits a $(111)$ peak from the TiC phase with a NaCl structure. The absence of other peaks from this phase suggests a significant $(111)$ texture. The position of the $(111)$-peak corresponds to a unit cell that is smaller than for pure TiC ($a=4.32$ \AA) and can be explained by a solid solution of Fe atoms into the carbide. From the Scherrer formula \cite{Cullity, Scardi} the grain size of the (Ti, Fe)C grains can be estimated to $\approx 100$ \AA.  The C1s XPS spectra shows only one strong peak at $\approx 281$ eV. This peak can be attributed to Me-C bonds. No indication of any significant amount of C-C bonds can be observed at $285$ eV. Consequently, in agreement with the results in Ref. \cite{Ola1}, the as-deposited films can be described as nanocrystalline films with (Ti,Fe)C grains without any free carbon matrix.  

A solid solution of Ti$_{0.6}$Fe$_{0.4}$C$_{0.5}$ is not thermodynamically stable.  Titanium carbide is known to exhibit a broad homogeneity range with carbon contents in the range from TiC$_{\approx 0.97}$ to TiC$_{0.5}$. However, the maximum solid solubility of Fe is less than 1 at$\%$ \cite{Vilars}. Annealing of the as-deposited Ti$_{1-\text{x}}$Fe$_{\text{x}}$C$_{1-y}$ film can therefore lead to two things: Initially small amounts of carbon can diffuse to the interface of the grains. The driving force for this is the rather weak Fe-C bonds in the carbide structure. The total energy of the system can be reduced by diffusion of carbon from the interior of the bulk to the surface under the formation of fairly strong C-C bonds. This effect has been experimentally demonstrated in Refs. \cite{Ola1,Ola3} and theoretically in a paper by R{\aa}sander {\it et al}.  \cite{rasanderunpubl}. In the present study the Ti$_{0.6}$Fe$_{0.4}$C$_{0.5}$ films have a very low carbon content close to the limits of the homogeneity range for the pure carbide. Thus, the carbon diffusion effect should be rather small. This was also confirmed by the C1s spectra in Fig. \ref{fig:Cs1}, showing no indication of C-C bonds after heat treatment. The second more important effect of annealing is that Fe should diffuse out from the grains and form metallic Fe precipitates. As can be seen in Fig.  \ref{fig:XRD}, clear XRD peaks from  $\alpha$-Fe is observed after 20 minutes. However, as can be seen in Fig.  \ref{fig:XRD} there is a gradual increase in the background intensity around $2\theta \approx 40 - 45^{\circ}$ during annealing. This can be attributed to the gradual  nucleation and precipitation of  Fe-grains. The exact details in phase formation from the nanocrystalline (Ti,Fe)C grains to a mixture of Ti-rich carbide and Fe is not known but as will be shown below this process may be very complex.

\begin{figure}
\begin{center}
\includegraphics[height=10cm]{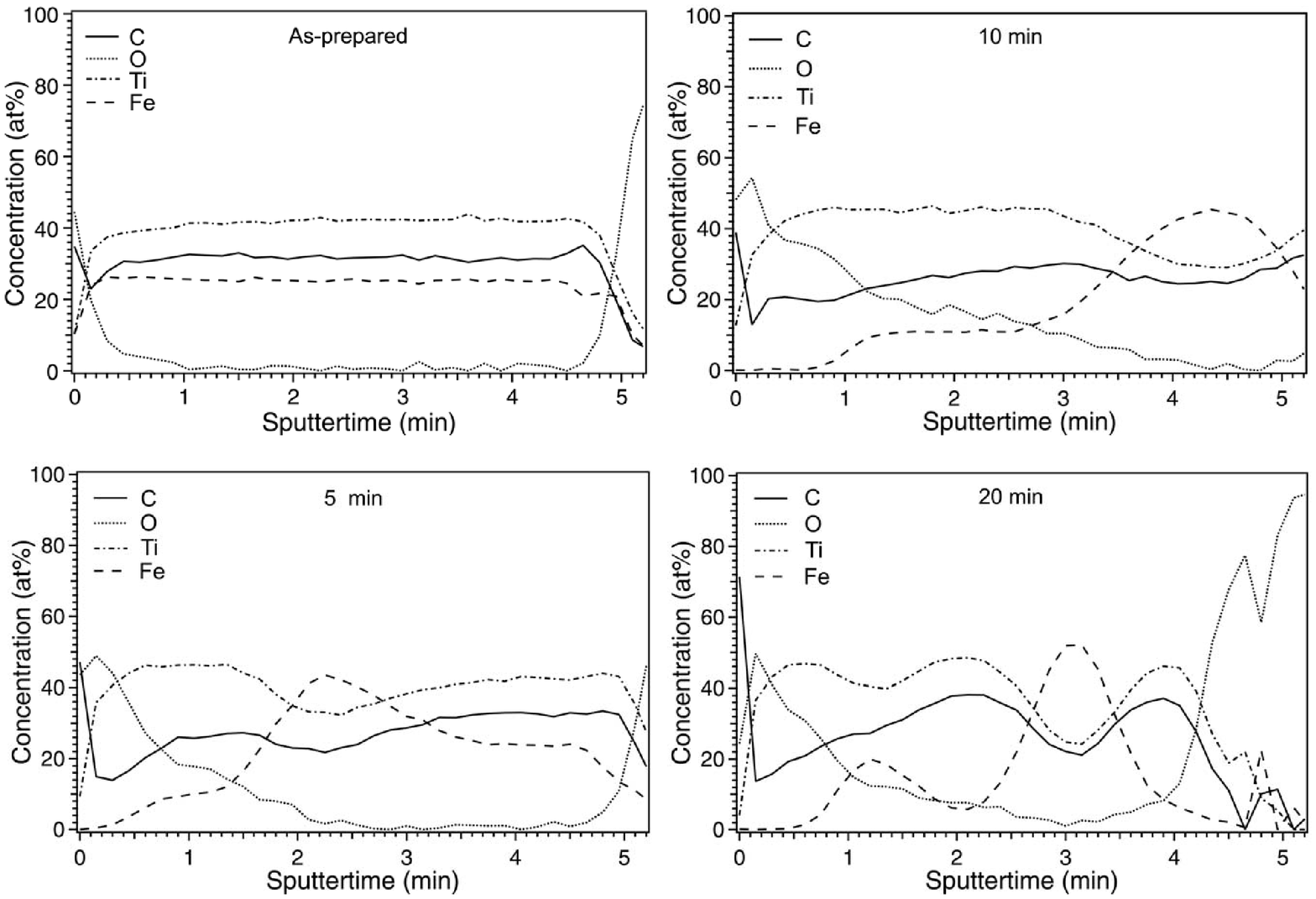}
\caption{XPS depth profiles vs. Ar$^{+}$ sputtering time  for the as-deposited film and for films after 5, 10  and 20 minutes minutes annealing at 650$^\text{o}$C. Annealing results in composition inhomogeneity and regions with different Fe-content.}
\label{fig:XPS}
\end{center}
\end{figure}

Figure \ref{fig:XPS} shows XPS-depth profiles of the Ti$_{1-\text{x}}$Fe$_{\text{x}}$C$_{1-\text{y}}$ films deposited on amorphous quartz. As can be seen, the atomic concentration of Ti, Fe and C in the as-deposited state is homogeneous, while the heat treatment induces a continuous change of the elemental distribution. It can be observed that with increasing heat treatment times there is a depletion of Fe close to the surface of the film, while enriched Fe-regions are simultaneously formed deeper into the film. As can be seen in Fig. \ref{fig:XPS}, heat treatments for 5 and 10 minutes, respectively, result in single regions enriched in Fe where the enriched region has moved deeper into the film after 10 minutes. For even longer heat treatment times two Fe-rich regions starts to develop. After 20 minutes of annealing two clear and well-defined regions of Fe is observed in the XPS-depth profile (cf. Fig. \ref{fig:XPS}). A slightly increased surface oxidation is also noted for the annealed samples. XPS results indicate that the oxygen is predominantly bonded to Ti. From the XRD-analysis it is also evident that the observed re-distribution and enrichment of Fe improves the crystallinity of the precipitating $\alpha$-Fe phase. As a matter of fact a clear Fe(110) diffraction peak is observed after 20 minutes of heat treatment (see Fig. \ref{fig:XRD}). 

\begin{figure}
\begin{center}
\includegraphics[height=3cm]{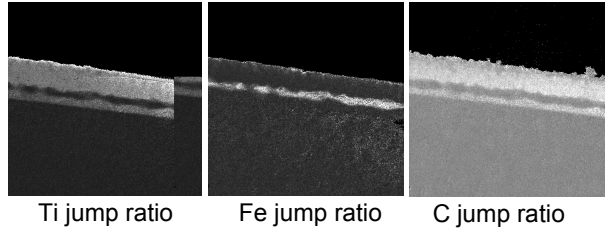}
\caption{Energy-filtered transmission electron microscope (EF-TEM) images of a (Ti,Fe)C film deposited on a single crystalline MgO(100) substrate and heat-treated for 60 minutes at 850$^\text{o}$C, showing regions with high concentrations of $\alpha$-Fe, Ti and C. Brighter contrast indicates higher element concentration.}
\label{fig:TEM}
\end{center}
\end{figure}

The XRD and XPS results presented above agrees very well with TEM-results obtained from studies on Ti$_{1-\text{x}}$Fe$_{\text{x}}$C$_{1-\text{y}}$ films deposited on single crystalline MgO(100) substrates \cite{unpublished}. These films have been grown epitaxially on the MgO(100) substrate and forms, in similarity with the results obtained for  amorphous quartz substrates in the present study, regions of $\alpha$-Fe upon vacuum-heat treatment. An illustrative example can be seen in Fig. \ref{fig:TEM} where cross-sectional energy-filtered TEM images of Ti, Fe and C are shown. The EF-TEM images clearly show that by annealing, the surface near region as well as the region close to the substrate is depleted in Fe and more similar to pure TiC and that an $\approx 100$ nm thick Fe rich layer is formed close to the substrate interface. Moreover, XRD results \cite{unpublished} as well as results from field dependent magnetization measurements (cf. Fig. \ref{fig:20min} (b)) indicate that the Fe layer is predominantly made up of $\alpha$-Fe.   

\subsection{\label{sec:Magn}Magnetic characterization}

\begin{figure}
\begin{center}
\includegraphics[height=7cm]{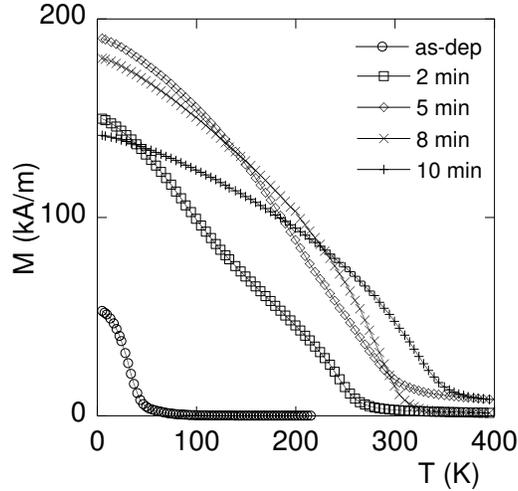}
\caption{Temperature dependence of the FC magnetization with an applied field of 50 Oe for different annealing times at at 650$^\text{o}$C. $T_{c}$ increases with annealing time.}
\label{fig:MvsT}
\end{center}
\end{figure}

Figure \ref{fig:MvsT} shows the temperature dependence of the magnetization in the temperature range from 5 K to 400 K for the as-deposited film and for the different annealing times ranging from 2 to 10 minutes. The as-deposited film exhibits a clear sign of magnetic ordering, with an ordering temperature of $\approx$ 50 K. As can be seen in the figure the magnetic behavior is strongly dependent on heat treatment time. Even for as small annealing time as 2 minutes, the low field magnetization is much larger in magnitude and more importantly there is a strong increase of the magnetic ordering temperature up to $\approx$ 250 K. Moreover, the heterogeneous composition of the annealed film, as seen in the previously discussed XPS results,  is reflected in the two-step-like magnetic transition, a feature which is also observed for 5 minutes of annealing. For 10 minutes of heat treatment, the magnetic ordering temperature has reached $\approx$ 350 K, and there is no clear sign of a two-step-like magnetic transition in the experimental temperature window. However, the low field magnetization remains comparably large in magnitude above the magnetic ordering temperature, which gives further support for the precipitation of nanocrystalline $\alpha$-Fe after sufficiently long annealing time. Thus, by carefully tuning the annealing time it can be claimed that the $T_{c}$-value for the Ti$_{1-\text{x}}$Fe$_{\text{x}}$C$_{1-\text{y}}$ substitutional solid solution can be increased up to approximately 350 K. 

\begin{figure}
\begin{center}
\includegraphics[height=14cm]{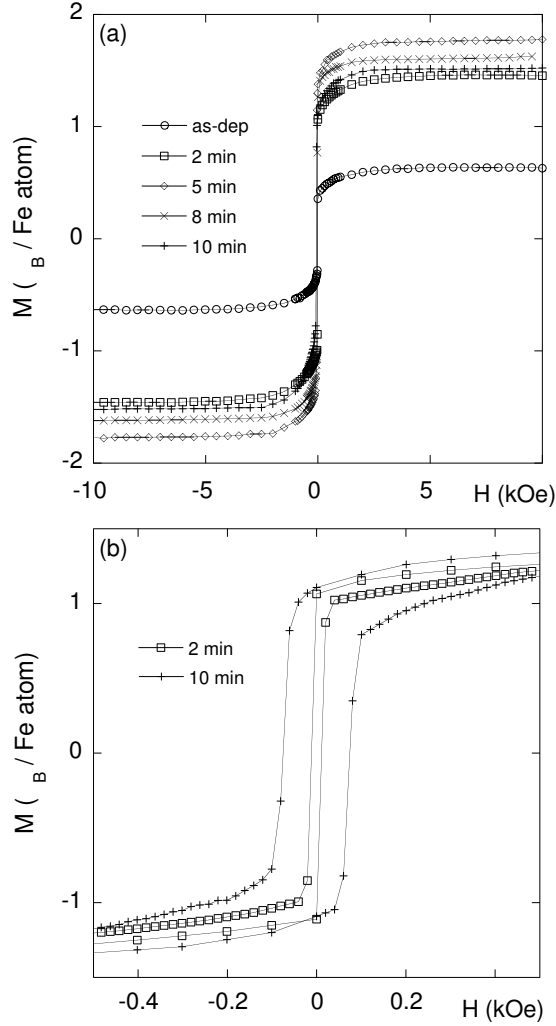}
\caption{a) Field dependence of the magnetization for different annealing times at 650$^\text{o}$C. b) The coercivity increases  due to formation of $\alpha$-Fe nanocrystals. }
\label{fig:MvsH}
\end{center}
\end{figure}

Figure \ref{fig:MvsH} (a) shows a comparison of the field dependence of the magnetization at 10 K for the as-deposited and annealed films. After heat treatment, the saturation magnetization ($M_{s}$) has increased for all films. This is explained by the composition inhomogeneity created by the heat treatment yielding regions with increased iron ($x$) and C-vacancy concentrations ($y$). The saturation magnetization is highest for the 5 minutes heat treatment film and slowly decreases as nanocrystalline $\alpha$-Fe starts to form. Figure \ref{fig:MvsH} (b) shows that the coercivity increases from less than 20 Oe for 2 minutes to 100 Oe for 10 minutes heat treatment indicating an increase in magnetic disorder as a result of the heterogeneous film composition, also evident from the x-ray diffractograms and from the XPS depth profiles.

\begin{figure}
\begin{center}
\includegraphics[height=13cm]{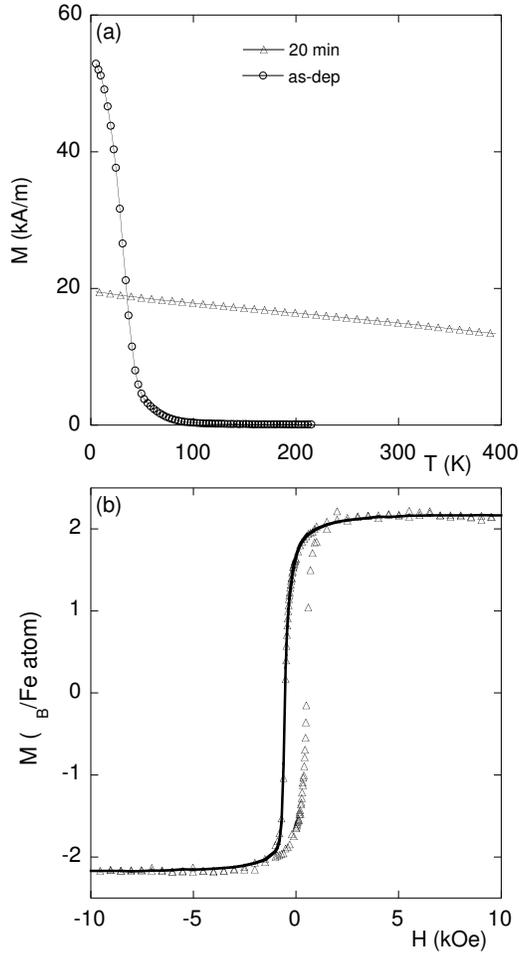}
\caption{a) Magnetization vs. temperature for the as-deposited film and the film annealed for 20 minutes at 650$^\text{o}$C. $T_{c}$ for the annealed film is far above 400 K. b) The coercivity for the annealed film increases significantly due to formation of $\alpha$-Fe nanocrystals (open triangles). Magnetization vs. field for the epitaxial film deposited on a single crystalline MgO substrate and annealed at 850$^\text{o}$C for 60 minutes is included for comparison (thick solid line), showing the magnetization values sweeping the field from $+$10 kOe to $-$10 kOe.}
\label{fig:20min}
\end{center}
\end{figure}

Figure \ref{fig:20min} (a) shows that after 20 minutes annealing the most of the Fe content has formed $\alpha$-Fe nanocrystals, since there is no sign of a magnetic ordering temperature in the experimental temperature window and $T_{c}$ for the $\alpha$-Fe phase is much higher than the upper boundary of the temperature window. Another confirmation of the $\alpha$-Fe phase is that the magnetic moment has increased to $\approx$ 2.1 $\mu_{B}/$Fe atom, as seen in Fig. \ref{fig:20min} (b). Moreover, the coercivity has increased to 500 Oe, which is not unexpected considering the nanocrystalline character of this phase and the shape anisotropy of the nanocrystals. For comparison, the magnetization vs. field results for the annealed epitaxial film has been included in the same figure, yielding the same magnetic moment ($\approx$ 2.1 $\mu_{B}/$Fe atom) at magnetic saturation. This result fortifies the interpretation of the EF-TEM images shown in Fig.  \ref{fig:TEM} that the Fe rich layer is predominantly made up of $\alpha$-Fe. It is interesting to note that both annealed films, i.e. the films deposited on amorphous quartz (annealed for 20 minutes at 650$^\text{o}$C) and single crystalline MgO (annealed for 60 minutes at 850$^\text{o}$C) substrates, exhibit almost identical coercive fields, which indicates that the Fe layers have very similar properties with respect to size, shape and packing of $\alpha$-Fe nanocrystals.   

\subsection{\label{sec:DFT}Theoretical results}

\begin{figure}
\begin{center}
\includegraphics[height=6cm]{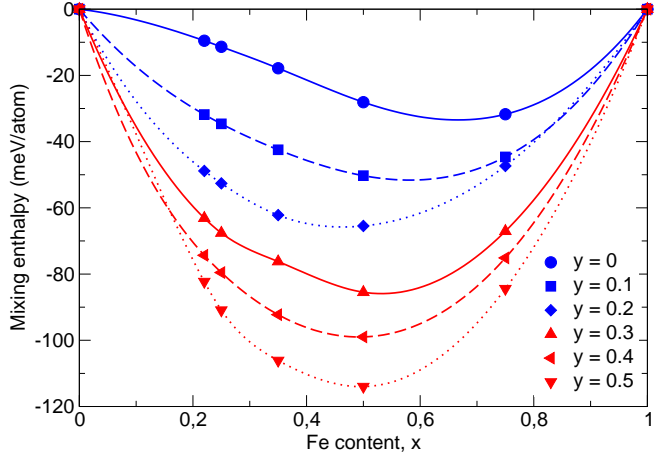}
\caption{(Color online) Calculated enthalpy of mixing for Ti$_{1-\text{x}}$Fe$_{\text{x}}$C$_{1-\text{y}}$ in the NaCl structure. For Fe concentration 0 $\leq$ x $\leq$ 1, and C vacancy concentration of 0 $\leq$ y $\leq$ 0.5. The lines connecting the dots in this figure is made by fitting a cubic spline to the calculated data.}
\label{fig:mixing}
\end{center}
\end{figure}
In Fig. \ref{fig:mixing} we show the calculated entalphy of mixing for Ti$_{1-\text{x}}$Fe$_{\text{x}}$C$_{1-\text{y}}$, evaluated via the equation
\begin{equation}\label{eq:mixing}
H_{mix} = E(\text{Ti}_{1-{\text{x}}}\text{Fe}_{\text{x}}\text{C}_{1-\text{y}})\\
-(1-\text{x})E(\text{TiC}_{1-\text{y}})-\text{x}E(\text{FeC}_{1-\text{y}}).
\end{equation}
We note that in Eq. (\ref{eq:mixing}) one of the reference energies is that of FeC$_{1-\text{y}}$ in the NaCl structure. This phase in not a stable phase, neither experimentally nor theoretically. Hence, one must be careful in interpreting the results of Eq. (\ref{eq:mixing}), which are shown in Fig. \ref{fig:mixing}. The stability criterion given by Eq. (\ref{eq:mixing}) simply answers if Fe and Ti mix on a metal carbide lattice in the NaCl structure.
If the mixing entalphy is negative there is a tendency for homogeneous mixing of the system, which is evidently the case for all concentrations, according to the result shown in Fig. \ref{fig:mixing}. Furthermore, the mixing increases with the amount of C vacancies, which means that the system has a larger tendency to mix as the C concentration is lowered. We also note that the curves show different behavior when varying the Fe concentration, x. For y = 0 the mixing enthalpy is lowest for large values for x with an energy minimum at about x = 0.75. When the C-vacancy concentration is increased the minimum enthalpy occurs at lower values for x. For y = 0.5 the minimum enthalpy occurs for x = 0.5, meaning that an equal mixture of Ti and Fe in (Ti,Fe)C is the most stable mixture.
\begin{figure}
\begin{center}
\includegraphics[height=6cm]{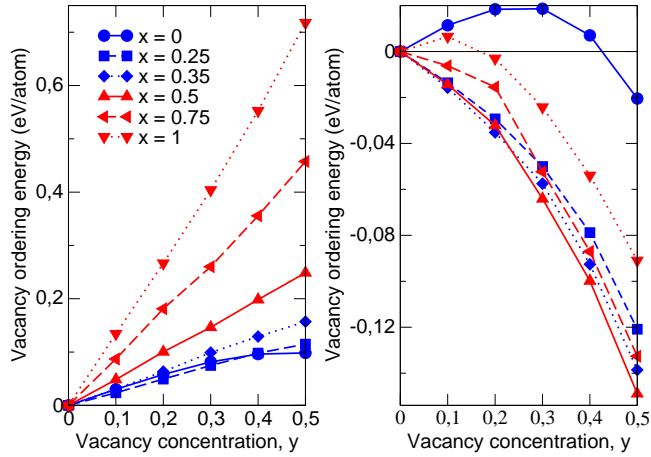}
\caption{(Color online) Calculated vacancy ordering energy for Ti$_{1-\text{x}}$Fe$_{\text{x}}$C$_{1-\text{y}}$. For Fe concentration 0 $\leq$ x $\leq$ 1, and C vacancy concentration of 0 $\leq$ y $\leq$ 0.5. The left panel is calculated according to Eq. (\ref{eq:vacord}) with the energy of the Ti$_{1-\text{x}}$Fe$_{\text{x}}$-phase at its equilibrium volume. The right panel is evaluated in a similar fashion, however, the volume  of the Ti$_{1-\text{x}}$Fe$_{\text{x}}$-phase is here identical to the volume of the carbide.}
\label{fig:vacancyord}
\end{center}
\end{figure}

In Fig. \ref{fig:vacancyord} we show the calculated vacancy ordering energy given by
\begin{equation}\label{eq:vacord}
E_{vac} = E(\text{Ti}_{1-{\text{x}}}\text{Fe}_{\text{x}}\text{C}_{1-\text{y}})\\
-(1-\text{y})E(\text{Ti}_{1-{\text{x}}}\text{Fe}_{\text{x}}\text{C})-\text{y}E(\text{Ti}_{1-{\text{x}}}\text{Fe}_{\text{x}}).
\end{equation}
If the vacancy ordering energy is negative it is more favorable for the C atoms to be randomly distributed on the C-lattice and if the ordering energy is positive it is more favorable for the C atoms to cluster, leaving other regions of the lattice empty of C. Eq. (\ref{eq:vacord}) can be evaluated in two ways, which are both displayed in Fig. \ref{fig:vacancyord}. Either the energy of the Ti$_{1-\text{x}}$Fe$_{\text{x}}$-phase is given at its equilibrium volume in which case we consider a local relaxation in the part of the system that is depleted of C (left panel of Fig. \ref{fig:vacancyord}), or it is given at the volume of the carbide where local relaxations are not considered (right panel of Fig. \ref{fig:vacancyord}).  From the right panel in Fig. \ref{fig:vacancyord} we see that C tends to cluster in TiC$_{1-\text{y}}$ for y $\leq$ 0.4 and for FeC$_{1-\text{y}}$ for y $\leq$ 0.2. In all other cases the ordering energy is negative which suggest a disordered distribution of the vacancies. However, if local changes in the volume are allowed we note that C-atoms tends to cluster for all values of x and y. Furthermore, we note that the ordering energy increases with increased Fe concentration. We believe that the results given in the left panel of Fig. \ref{fig:vacancyord} are the most realistic vacancy ordering energies, with which one should compare the experimental results to.
\begin{figure}
\begin{center}
\includegraphics[height=5cm]{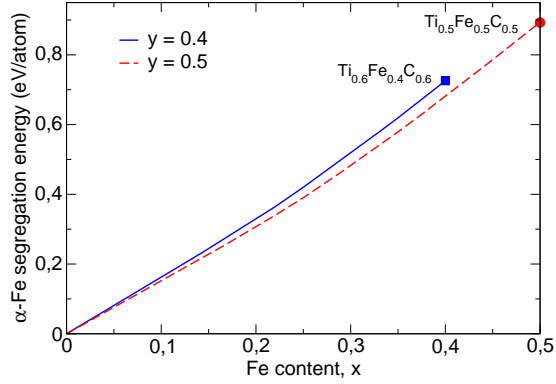}
\caption{(Color online) Energy gained by releasing Fe from Ti$_{1-\text{x}}$Fe$_{\text{x}}$C$_{1-\text{y}}$ in the form of $\alpha$-Fe for y = 0.4 and 0.5. Obtained according to Eq. (\ref{eq:Feseg}).}
\label{fig:segregation}
\end{center}
\end{figure}

We continue our discussion of the energetics of Ti$_{1-\text{}}$Fe$_{\text{x}}$C$_{1-\text{y}}$ by analyzing the energy for $\alpha$-Fe segregation which is evaluated according to
\begin{equation}\label{eq:Feseg}
E_{s}(\alpha\text{-Fe}) = E(\text{Ti}_{1-{\text{x}}}\text{Fe}_{\text{x}}\text{C}_{1-\text{y}})\\
-(1-\text{x})E(\text{Ti}\text{C}_{\text{z}})-\text{x}E(\alpha \text{-Fe})
\end{equation}
where z = $\frac{1-\text{y}}{1-\text{x}}$. The results of Eq. (\ref{eq:Feseg}) are given in Fig. \ref{fig:segregation} for two values of y (y = 0.4 and y = 0.5). It is clear that by increasing the Fe concentration it is energetically more favorable for Fe to segregate from the carbide and form $\alpha$-Fe. This holds for all considered values of x and y, and demonstrates that the films considered experimentally here are all metastable. The results of Fig. \ref{fig:segregation} are consistent with the low solubility of Fe in bulk TiC.

\begin{figure}[htbp]
\begin{center}
\includegraphics[height=6cm]{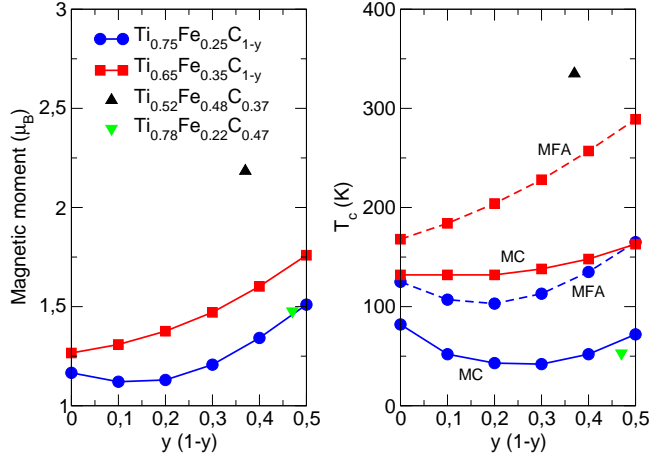}
\caption{(Color online) Calculated magnetic moments (left panel) and critical temperatures (right panel) from both mean field and Monte Carlo simulations as a function of carbon vacancy concentration, y, for different Fe concentrations, x. The MC data for the critical temperatures are drawn with a solid line while the corresponding MFA values are drawn with a dashed line. For the specified compositions Ti$_{0.52}$Fe$_{0.48}$C$_{0.37}$ and Ti$_{0.78}$Fe$_{0.22}$C$_{0.47}$ only MC values are given.}
\label{fig:tc}
\end{center}
\end{figure}

In Fig. \ref{fig:tc} we show the calculated magnetic moments for the Fe atoms in Ti$_{0.75}$Fe$_{0.25}$C$_{1-\text{y}}$ and Ti$_{0.65}$Fe$_{0.35}$C$_{1-\text{y}}$  along with the calculated critical temperatures from both MC and MFA. We also show calculated data points for two specific compositions, namely Ti$_{0.52}$Fe$_{0.48}$C$_{0.37}$ and Ti$_{0.78}$Fe$_{0.22}$C$_{0.47}$,  corresponding to regions of high and low Fe concentrations (cf. Fig. \ref{fig:XPS}), respectively. The general trend is that the magnetic moment and ordering temperature both increase with increasing Fe content. The same is true when increasing the C vacancy concentration although the effect is much stronger when increasing the Fe content.
We also note that the MFA gives larger values than MC for the ordering temperatures.
Focusing on the curve for  Ti$_{0.65}$Fe$_{0.35}$C$_{1-\text{y}}$, which is the one closest resembling our experimental sample, we see that the magnetic moments and MFA ordering temperatures have a similar increasing behavior as a function of C vacancy concentration, while the MC ordering temperature stay more or less constant at about 130 K to 135 K for low vacancy concentrations with an increase to 165 K for y = 0.5. 
Furthermore, there is an initial descent in the value for the magnetic moment and ordering temperature for x=0.25, which will become apparent when considering the behavior of the magnetic exchange parameters. 
However, it is still valid to conclude that by increasing the Fe content or the C vacancy concentration the magnetic moment as well as the ordering temperature increases.

By comparing the calculated magnetic moments and ordering temperatures to the experimentally determined values we note that upon annealing between 2 and 10 minutes the magnetic moment of the Fe atoms varies between 1.43 $\mu_{B}$ to 1.82 $\mu_{B}$ and the ordering temperature varies from  270 K to 360 K. The theoretical results obtained for x = 0.35 and y = 0.5 give a moment of 1.76 $\mu_{B}$ and an ordering temperature of 165 K. The calculated moment is within the experimental interval but the calculated ordering temperature is somewhat too low. However, since the annealed samples show regions of different compositions, as shown by the XPS depth profiles, this has to be taken into consideration. The theoretically considered compositions of high (x = 0.48, y = 0.63) and low (x = 0.22, y = 0.53) Fe concentrations correspond to regions found in the sample after 5 minutes of annealing. Here the experimentally obtained ordering temperature is 310 K. The calculated ordering temperatures are 340 K and 50 K for the Fe-rich and Fe-poor compositions respectively. It has been shown \cite{Skomski} that for magnetic composites the ordering temperature is close to the highest value of the materials forming the composite, i.e. 340 K in our case, and bearing this in mind one can conclude that there is good agreement between the experimentally obtained ordering temperature and the theoretical value.

The experimentally obtained magnetic moment is 1.82 $\mu_{B}$, while the theoretical moments are 2.18 $\mu_{B}$ (Fe-rich) and 1.48 $\mu_{B}$ (Fe-poor). Considering that the experimentally obtained moment is an averaged quantity (total magnetization divided by the number of Fe atoms) over the whole sample, we compare this value with the average between the theoretical values for Fe-rich and Fe-poor regions, assuming equal volumes for the two regions, given by $\bar{\mu} = (\mu_{\text{Fe-rich}}+\mu_{\text{Fe-poor}})/2$ = 1.83 $\mu_{B}$. This averaged moment is in good agreement with the experimentally obtained moment. According to the above discussion we conclude that the theoretically obtained magnetic moments and ordering temperatures are in good agreement with experiment when considering that the experimental sample consist of regions of  both Fe-rich and Fe-poor regions. 

\begin{figure}[htbp]
\begin{center}
\subfigure[]{\includegraphics[height=5cm]{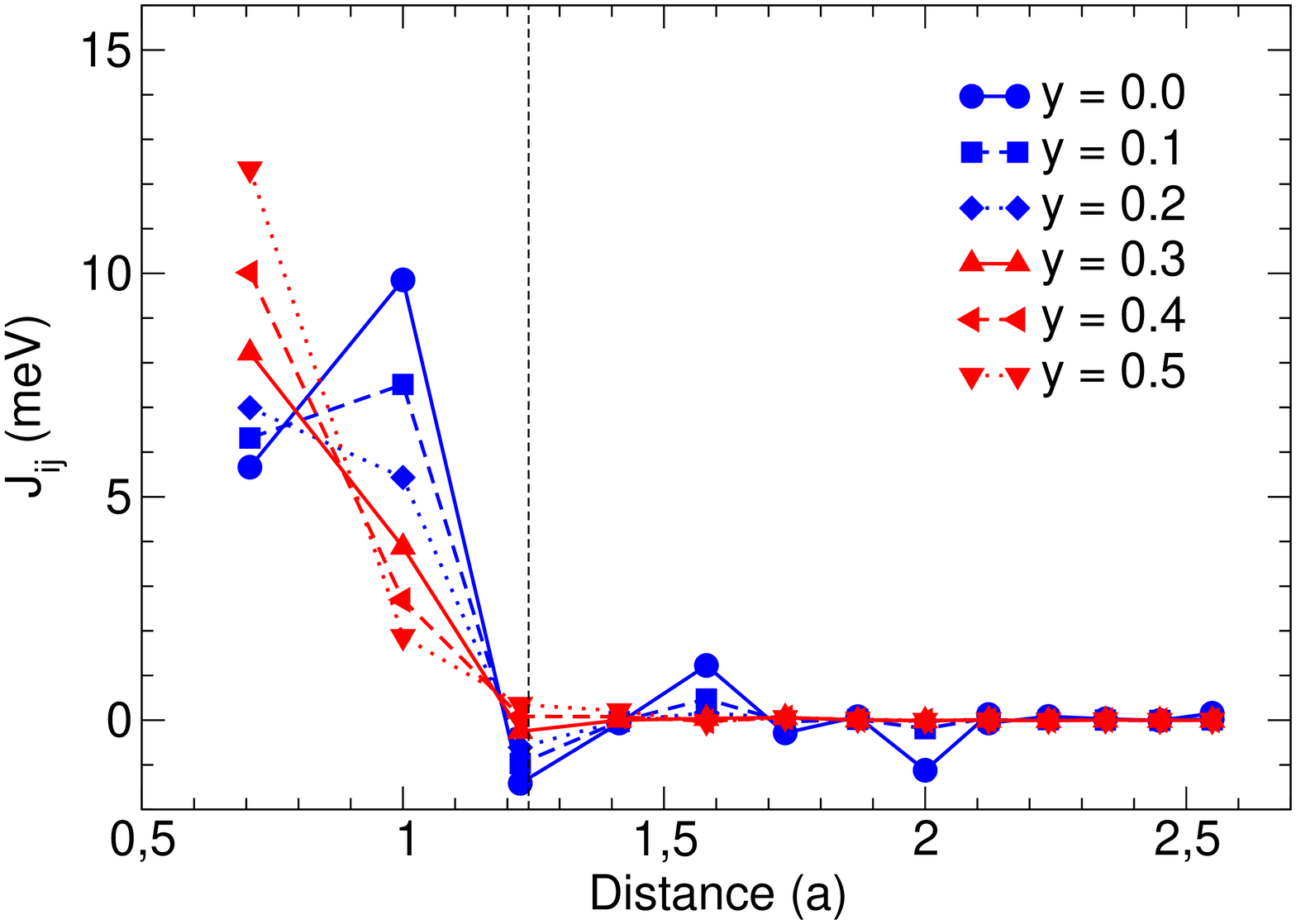}}
\subfigure[]{\includegraphics[height=5cm]{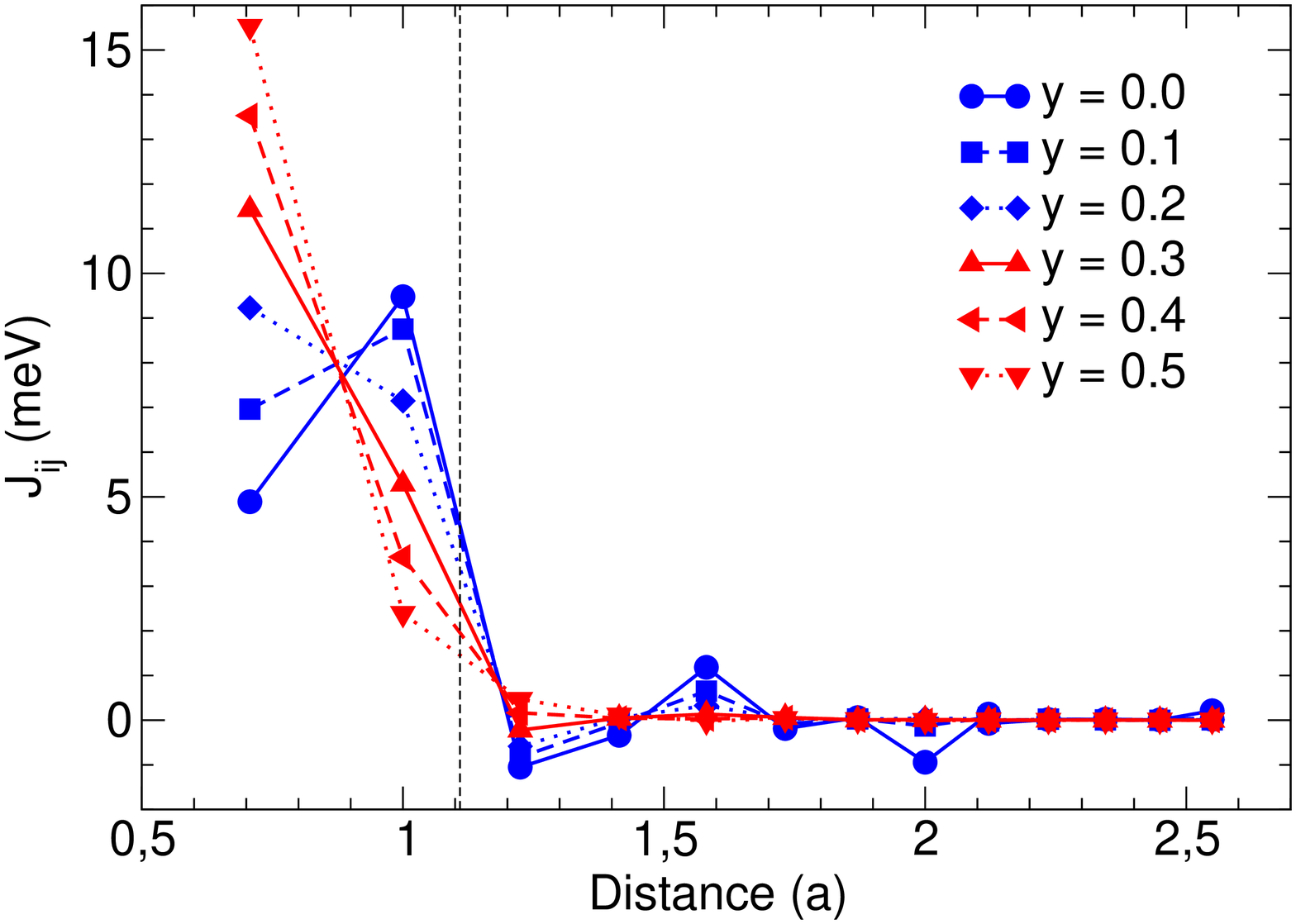}}
\caption{(Color online) Calculated exchange parameters for Ti$_{1-\text{x}}$Fe$_{\text{x}}$C$_{1-\text{y}}$ with x = 0.25 (a) and $0.35$ (b). Note the oscillating behavior for low vacancy concentration, where the strength of the nearest neighbor interaction is lower than the second neighbor interaction for y = 0 and 0.1, while the exchange interaction is rapidly decreasing for the higher vacancy concentrations. The dotted vertical line indicates the average distance between Fe atoms for the corresponding concentrations.}
\label{fig:exchange}
\end{center}
\end{figure}

The calculated magnetic exchange parameters, which were used in the evaluation of the ordering temperatures, are presented in Fig. \ref{fig:exchange}. The most apparent fact is that the exchange parameters for low carbon vacancy concentrations (y = 0, 0.1 and 0.2) oscillate between positive and negative values, meaning that the interaction is ferro- or antiferromagnetic depending on the distance between magnetic atoms. For even greater C vacancy concentrations the interaction is rapidly decreasing with increasing distance between magnetic atoms. Furthermore, the nearest neighbor interaction is lower than the second nearest neighbor interaction for low  C-vacancy concentration, while there is a change in behavior for $\text{y}\geq 0.2$, where the nearest neighbor interaction is stronger.
This is reasonable considering the B1 crystal structure and the chemical bonding of TiC. The nearest neighbor Fe-Fe interaction distance is given by $a/\sqrt{2}$, where $a$ is the lattice parameter of the primitive cell, while the second nearest neighbor is at a distance $a$ away. If the system would have been a binary TiFe alloy on a regular face centered cubic lattice there would have been a considerable $d$-orbital overlap between nearest neighbor metal atoms. However, this is not the case in ternary Ti$_{1-\text{x}}$Fe$_{\text{x}}$C$_{1-\text{y}}$, since the inclusion of a carbon lattice has increased the volume of the system resulting in a reduced $d$-band overlap between the atoms on the metal lattice \cite{Moruzzi}.
The chemical bonding in TiC is primarily due to strong covalent bonds formed by Ti-$d$ and C-$p$ electrons separated by a distance $a/2$ from each other. The second nearest neighbor interaction for a Fe atom in Ti$_{1-\text{x}}$Fe$_{\text{x}}$C$_{1-\text{y}}$ is along the direction of the covalent bond, yielding a stronger magnetic interaction due to the significant orbital overlap along this direction. However, if there is a substantial number of C-vacancies the covalent bonding between metal atoms and carbon will give way for a larger $d$-band overlap between neighboring metal atoms, therefore changing the relative strength of the magnetic interaction between first and second shell neighbors, which can readily be observed in Fig. \ref{fig:exchange}.

Furthermore, by comparing the exchange parameters for x = 0.25 and x = 0.35 we conclude that the exchange parameters are generally stronger for x = 0.35 which is shown most prominently in the first neighbor interaction in Fig. \ref{fig:exchange}, especially for high vacancy concentrations. This is the reason for the  higher ordering temperatures shown in Fig. \ref{fig:tc} for the phases with a higher Fe concentration. Figure \ref{fig:exchange} also shows that the average distance between Fe atoms in the system lies in regions of different behavior of the exchange parameters. For x = 0.25 the average distance lies in a region of anti-ferromagnetic coupling between Fe atoms for $\text{y}\leq 0.2$ and a very low ferromagnetic coupling for $\text{y}\geq 0.2$. When the Fe concentration is increased to x = 0.35 the averaged distance between Fe atoms are in a position of ferromagnetic coupling for all y within the calculated range. It is therefore possible to conclude that the ferromagnetic properties of x = 0.35 is greater than for x = 0.25 which is readily seen in Fig. \ref{fig:tc}.
From the exchange parameters for x = 0.25 we also note that the next nearest neighbor interaction goes down faster than the increase in the nearest neighbor interaction for $\text{y}\leq 0.2$. This is the reason for the initial descent in the MFA ordering temperature for x = 0.25.

\begin{figure}[tbp]
\begin{center}
\subfigure[]{\includegraphics[height=5cm]{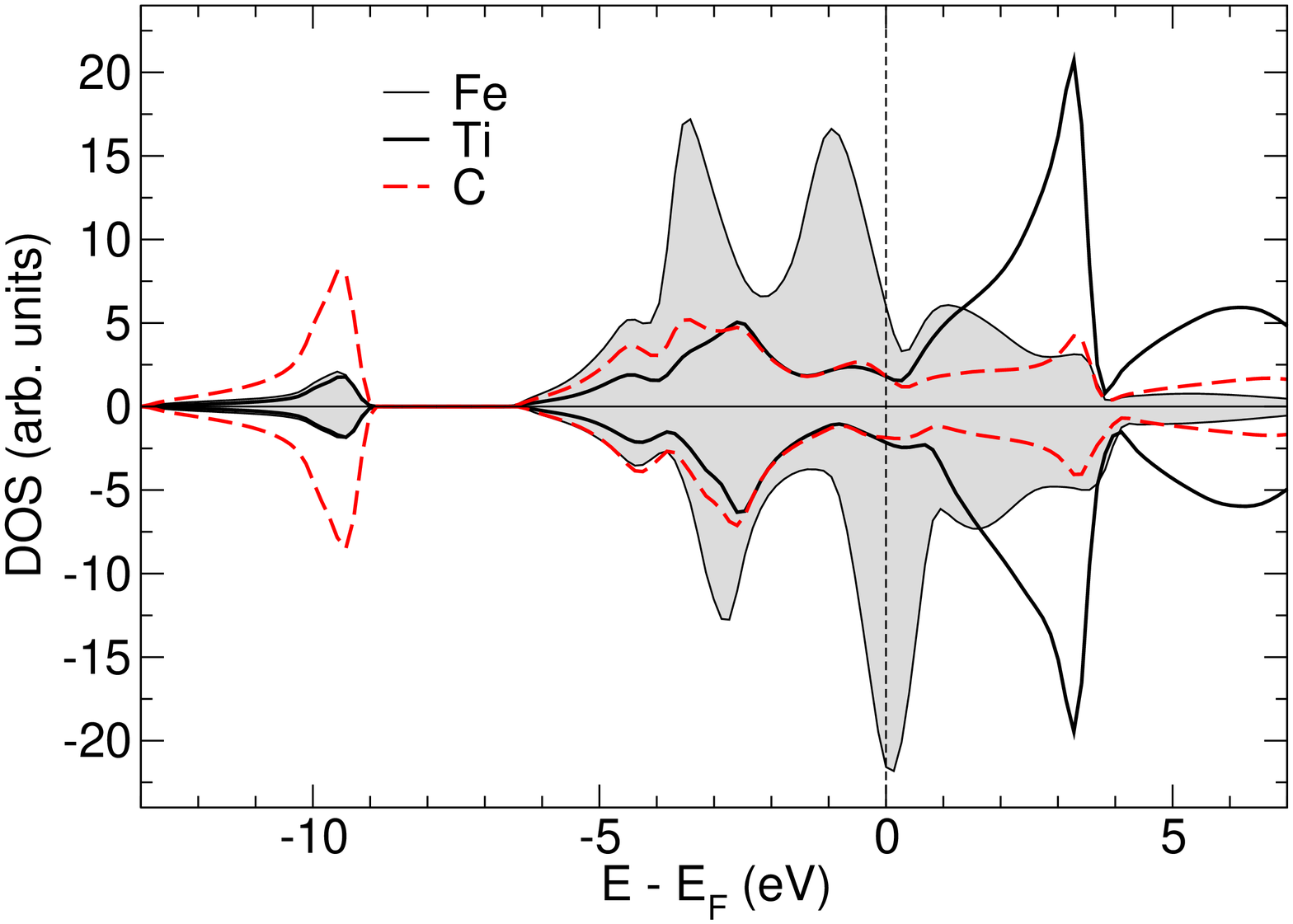}}
\subfigure[]{\includegraphics[height=5cm]{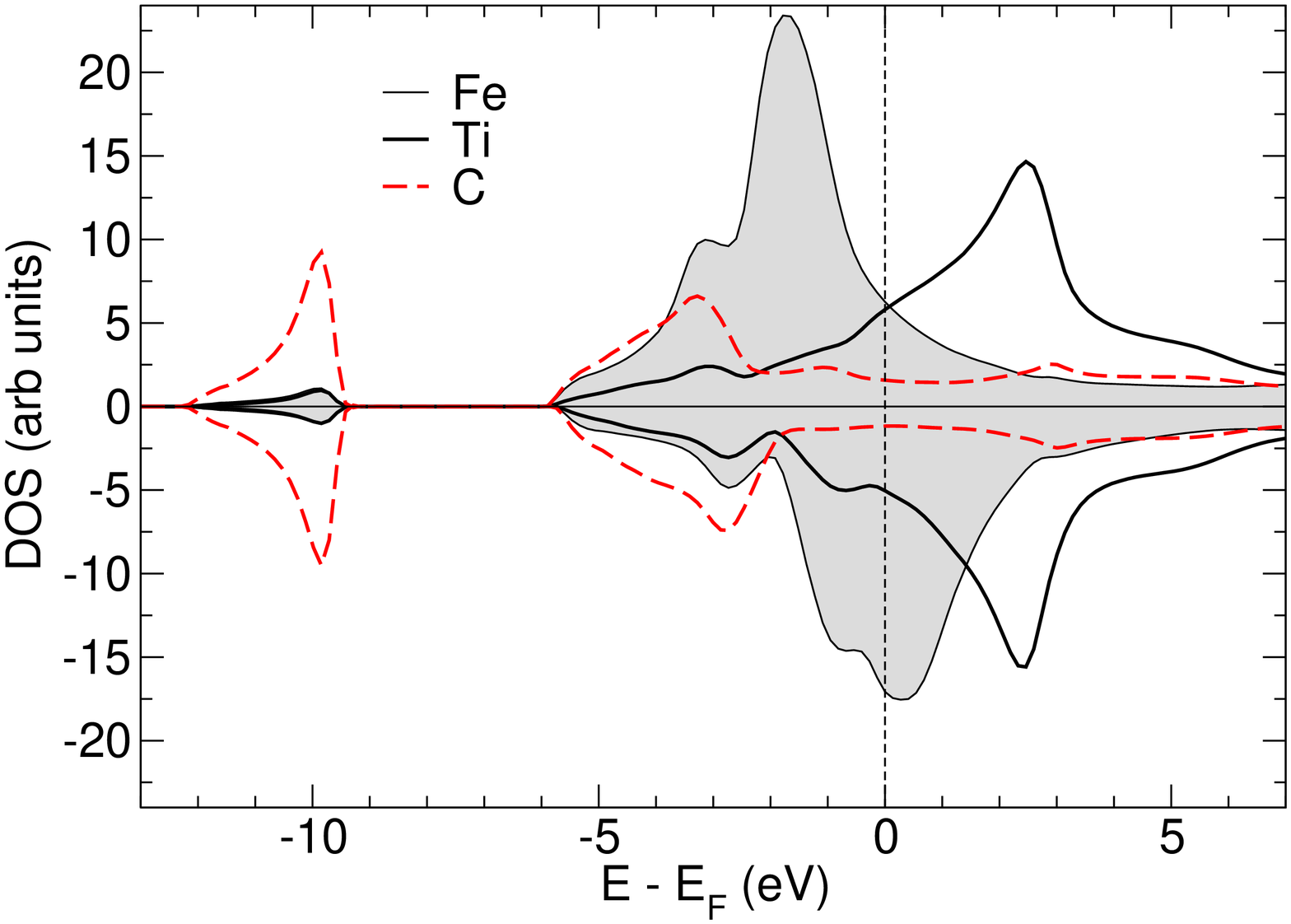}}
\caption{(Color online) Projected density of states (DOS) for Ti$_{0.65}$Fe$_{0.35}$C$_{1-\text{y}}$ for y = 0 (a) and 0.5 (b). The Fermi level is indicated by a solid vertical line and the scale on the y axis is the same for each of the DOS. The DOS for Fe are filled with grey color for clarity.}
\label{fig:dos}
\end{center}
\end{figure}

In Fig. \ref{fig:dos} we show the calculated density of states for each atom type in Ti$_{0.65}$Fe$_{0.35}$C$_{1-\text{y}}$, for two different vacancy concentrations (y = 0 and 0.5). When there are no C-vacancies present there are two distinct peaks in the DOS for Fe, both in the majority and minority spin-channels. By increasing the number of C-vacancies these peaks move closer together and furthermore the middle of each Fe band has moved to lower (majority-spin) and higher (minority-spin) energies. This increased shift in the band centers for majority and minority spin-channels is responsible for the increased magnetic moment when decreasing the amount of C in the system.
At the same time the Ti DOS has become broader and moved toward lower energies for both up and down spin-channels. This signifies a more pronounced $d$-orbital overlap between Ti atoms for higher C-vacancy concentrations, consistent with the previous discussion for the exchange parameters. 
The DOS at about -10 eV which is dominated by C 2s-states has become narrower and with a lower DOS for Ti and Fe states in this region which further supports the argument of less pronounced bonding between metal atoms and C.

\section{Summary and Conclusions}

Thin films of Ti$_{1-\text{x}}$Fe$_{\text{x}}$C$_{1-\text{y}}$ have been demonstrated to be obtained by magnetron sputtering. In these films, a solid substitutional solution is formed where Fe replaces Ti in the TiC structure. The films can be deposited in a wide range of compositions with respect to both carbon and metal content. At high Fe-contents, metallic $\alpha$-Fe forms precipitations. $\alpha$-Fe formation as nano-crystallites is also favored by a long deposition time and by a post-annealing at 650$^{o}$C to 850$^{o}$C. TEM studies on epitaxial samples on single crystalline MgO substrates (Fig. \ref{fig:TEM}) suggest that, by annealing, the surface near region as well as the region close to the substrate is depleted in Fe and more similar to pure TiC. 

The results are further supported by XPS studies (Fig. \ref{fig:XPS}) where it can be seen how the element composition changes as a result of heat treatment. All elements seemingly redistribute during heat treatment and with enough annealing an $\alpha$-Fe layer, embedded in TiC depleted in Fe, is formed close to the substrate surface (cf. Fig. \ref{fig:XPS}). These experimental findings are in agreement with our first principles theory, which show that Fe doped in an TiC film becomes a meta-stable system. However, theory shows that if the formation of $\alpha$-Fe is blocked, the Fe atoms are randomly distributed on the metal sites of the TiC lattice. The experimental results indicate that the formation of $\alpha$-Fe is blocked during the early stages of annealing, but that the formation of nanocrystalline $\alpha$-Fe will prevail after long enough annealing time. The first principles calculations also show that if vacancies are introduced on the C sublattice they tend to cluster, so that one part of the sample is richer in C and another part has a lower C concentration.

The magnetic properties of the Fe doped TiC films have also been investigated experimentally and theoretically. We show that the magnetic moment and ordering temperature are rather substantial, at least for the concentrations of Fe considered here. The interatomic exchange interactions have also been analyzed and found to reflect the underlying chemical bonding of the material. The trend of $\alpha$-Fe to form as a separate layer inside the TiC matrix, is an interesting avenue to grow magnetic tri- or multilayers by a repetition of the growth procedure presented here and subsequent heat treatment. The results of such a study are outside of the scope of the present article.


\section{Acknowledgements}
We acknowledge support from the Swedish Research Council, Swedish Foundation for Strategic Research and the Swedish National Allocations Committee.


\end{document}